\documentclass[twoside]{article}
\usepackage{fleqn,espcrc2}
\usepackage{graphicx}

\title{ Monte Carlo Study of Two-Color QCD with Finite Chemical Potential \\
-- Status report of Wilson fermion simulation -- 
 }

\author{
     S. Muroya\address{Tokuyama Women's College, Tokuyama, 745-8511, Japan},
     A. Nakamura\address{RIISE, Hiroshima University, 
                         Higashi-Hiroshima 739-8521, Japan}
     and
     C.Nonaka\address{Department of Physics, Hiroshima University,
                             Higashi-Hiroshima 739-8526, Japan },
}
       
\begin{document}

\begin{abstract}
Using Wilson fermions, we study SU(2) lattice QCD 
with the chemical potential at $\beta=1.6$.
The ratio of fermion determinants is evaluated at each Metropolis
link update step.
We calculate the baryon number density, the Polyakov loops and the
pseudoscalar and vector masses on $4^4$ and $4^3\times 8$ lattices.  
Preliminary data show the pseudoscalar meson becomes massive around
$\mu=0.4$, which indicates the chiral symmetry restoration. 
The calculation is broken down when approaching to the transition
region.
We analyze the behavior of the fermion determinant and  
eigen value distributions of the determinant, which shows a peculiar
``Shell-and-Bean'' pattern near the transition.

\vspace{1pc}
\end{abstract}

% typeset front matter (including abstract)
\maketitle

\section{INTRODUCTION}

Lattice study of QCD has been expected to provide useful
informations to understand non-perturbative aspects of
quark/gluon physics.
Especially at finite temperature, it predicts the 
confinement/deconfinment transition and is able to describe many 
features of hadrons and quark gluon plasma (QGP).  Lattice QCD 
offers a sound base of QGP physics, which has become very 
important issue of physics because of currently going 
active experiments at CERN SPS and BNL RHIC. See \cite{Satz}.

Contrary to the finite temperature calculation, the progress in
lattice QCD study of the finite density has been rather slow.
This is because of the well known complex action problem.
Indeed after the first QCD dynamical quark simulation with the
chemical potential was done for SU(2) color group \cite{Nakamura84},
to our knowledge, no full SU(3) QCD calculations had been tried.
A trial to put the phase coming from the determinant into observables
suffers from large fluctuation even at $4^4$ size lattice near the
phase transition \cite{Gocksch,Nakamura90,Toussaint}.
Stephanov shows that the quench approximation is not the correct
$N_f=0$ limit of full QCD \cite{Stephanov}.
We still wait for good news concerning Glasgow method (see a good
review by Barbour \cite{Barbour98} and references therein), and 
finite density method \cite{kaczmarek}.

Recently the situation has been changed; 
Due to the progress in analytical investigations \cite{Kogut00,Son}, 
we have a hope to obtain informations on concerning QCD by
studying QCD-like theories such as SU(2) QCD, models with
quarks in the adjoint representation and QCD at finite isospin density;
they are expected to have less difficulties in numerical analyses.
In these years, there are indeed high activities in Monte Carlo 
calculations with dynamical quark of these 
models \cite{Lombardo99a,Lombardo99,Morrison,Montvay,Sinclair}.

In this paper, we report our recent work on SU(2) QCD with
Wilson fermions to study finite density states.

\section{ALGORITHM}

The chemical potential, $\mu$, is introduced in the fermion action,
$\bar{\psi} W \psi$, as
\begin{eqnarray}
\lefteqn{W(x,x') = \delta_{x,x'}}\hspace{12cm} \nonumber\\ 
\lefteqn{ - \kappa \sum_{i=1}^{3} \left\{ 
        (1-\gamma_i) U_i(x) \delta_{x',x+\hat{i}} 
\right. 
\left.
      + (1+\gamma_i) U_i^{\dagger}(x') \delta_{x',x-\hat{i}} \right\}}
\hspace{12cm}
\nonumber \\
\lefteqn{      - \kappa \left\{ 
        e^{+\mu a}(1-\gamma_4) U_4(x) \delta_{x',x+\hat{4}} 
\right. }\hspace{12cm}
\nonumber \\
\lefteqn{
\left.
      + e^{-\mu a}(1+\gamma_4) U_4^{\dagger}(x') \delta_{x',x-\hat{4}} \right\}
 (1)}\hspace{9cm}
\label{Wfermion}
\end{eqnarray} \noindent
by Hasenfratz and Karsch \cite{HK84} to avoid an infinity 
in the energy density.  
As the lattice spacing $a$ tends to zero, Eq.(\ref{Wfermion}) gives
$W(\mu) = W(0) + \kappa \mu a \bar{\psi}\gamma_4\psi + O(a^2)$.
The above formula was independently obtained in Ref.\cite{Nakamura85}
by the following naive argument:
In the continuum perturbation, the chemical potential is introduced
by the substitution $p_4 \rightarrow p_4 - i\mu$ in fermion propagators.  
In order to have this continuum limit, the lattice fermion propagator 
should have the form,
\begin{eqnarray}
1/
\lefteqn{ \Bigl( 1
 - \kappa \sum_{i=1}^{3} \left\{
   (1-\gamma_i) e^{i p_i a} + (1+\gamma_i) e^{-i p_i a} \right\} 
\Bigr. }\hspace{7.0cm}
\nonumber \\
\Bigl.
\lefteqn{   - \kappa \left\{
   (1-\gamma_4) e^{i (p_4-i\mu) a} + (1+\gamma_4) e^{-i (p_4-i\mu) a} 
   \right\} \Bigr )}\hspace{7.0cm}
\label{FreeWfermion}
\end{eqnarray}

Little is known about the behavior of dynamical fermion simulations
when the chemical potential is introduced.  We therefore decide to
employ an algorithm where the ratio of the determinant,
\begin{equation}
 \frac{ \mbox{det} W(U+\Delta U) }{ \mbox{det} W(U) } 
= \mbox{det} (I + W(U)^{-1} \Delta W)
\end{equation} \noindent
is evaluated explicitly at each Metropolis update process, 
$U \rightarrow U+\Delta U$, 
where 
$\Delta W \equiv W(U+\Delta U) - W(U)$ \cite{Barbour87,Nakamura88,PDF87}.  
An essential ingredient of the algorithm is
the following Woodbury formula, 
\begin{eqnarray}
\lefteqn{(W+\Delta W)^{-1}} \hspace{6cm}
\nonumber \\
\lefteqn{ = W^{-1}- W^{-1}\Delta W
\left( I + W^{-1}\Delta W\right)^{-1} W^{-1}}
\hspace{5.7cm}
\label{Woodbury}
\end{eqnarray}

Suppose we update link variables $U_\mu(x)$'s only on
a subset $H$ of whole lattice.  Then $\Delta W \neq 0$
only on $H$.  Woodbury formula (\ref{Woodbury}) holds even
if the matrix space is limited on $H$.  In this case we can get
the ratio of the fermion determinant as far as $U_\mu(x)$'s
are updated inside $H$.  We take a $2^4$ hypercube as $H$.
When we go to the next hypercube, $(W^{-1})_H$'s
are initialized by CG method.

\section{RESULTS}

\begin{figure}
\begin{center}
\includegraphics[width= 0.9\linewidth]{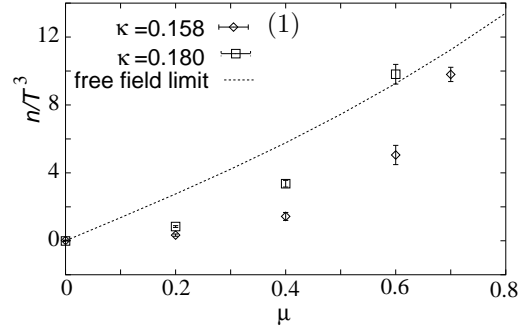}
\end{center}
\caption{Baryon number density divided by $T^3$ as
a function of $\mu$.}
\label{Fig-density}
\end{figure}

First we calculate the expectation value
of the density,
\begin{equation}
<n> = \frac{1}{\beta V_s} \frac{\partial}{\partial\mu} \log Z
\end{equation} \noindent
where $V_s$ is the spatial volume $N_xN_yN_z$.
In Fig.\ref{Fig-density},  we plot $<n>/T^3$, as a function of $\mu$,
which is dimensionless.
The dotted line corresponds to the free quark case
obtained by setting $U_\mu(x)=1$ and $\kappa=1/8$.
When $\kappa$ becomes large, the density reaches to
the free case quickly.

\begin{figure}
\begin{center}
\includegraphics[width= \linewidth]{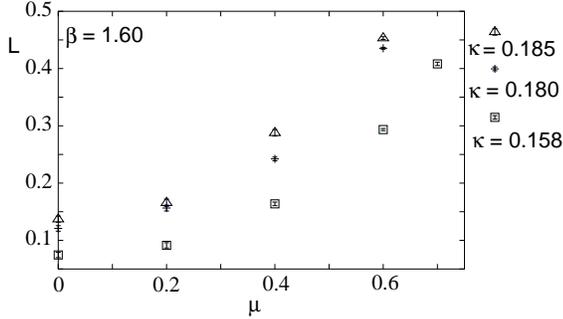}
\end{center}
\caption{Polyakov loop expectation value as a function of 
$\mu$ for $\kappa$=0.158, 0.180 and 0.185.}
\label{Fig-Polyakov}
\end{figure}

The Plyakov line $<L>$ also increases as a function of $\mu$
as shown in Fig.\ref{Fig-Polyakov}.  When the hopping parameter,
$\kappa$, becomes large from 0.158 to 0.185, 
values of Polyakov line increase.
Since the lattice size is small, 
no sharp increase of $<L>$ is seen, but
large values of $<L>$ indicate quarks become free from the
confinement force at $\mu> 0.4$.

%%%%%%%%%%%%%%%%%%%%%%%%%%%%%%%%%%%%%%%%%%%%%%%%%%%%%
% Density 
%
%%%%%%%%%%%%%%%%%%%%%%%%%%%%%%%%%%%%%%%%%%%%%%%%%%%%%
% Plyakov loop 
%
%%%%%%%%%%%%%%%%%%%%%%%%%%%%%%%%%%%%%%%%%%%%%%%%%%%
% ratio of W  \kappa = 0.156
% \begin{figure}
% \begin{minipage}{0.5\linewidth}
% \includegraphics[width= \linewidth]{pdetw1.eps}
% \caption{det $W$}
% \end{minipage}
% \hspace{0.5cm}
% \begin{minipage}{0.5\linewidth}
% \includegraphics[width= \linewidth]{pdetw4.eps}
% \caption{det $W$}
% \end{minipage}
% \end{figure}
%%%%%%%%%%%%%%%%%%%%%%%%%%%%%%%%%%%%%%%%%%%%%%%%%%%
% ratio of W \kappa = 0.180
\begin{figure}[htb]
\begin{minipage}{0.45\linewidth}
\includegraphics[width= \linewidth]{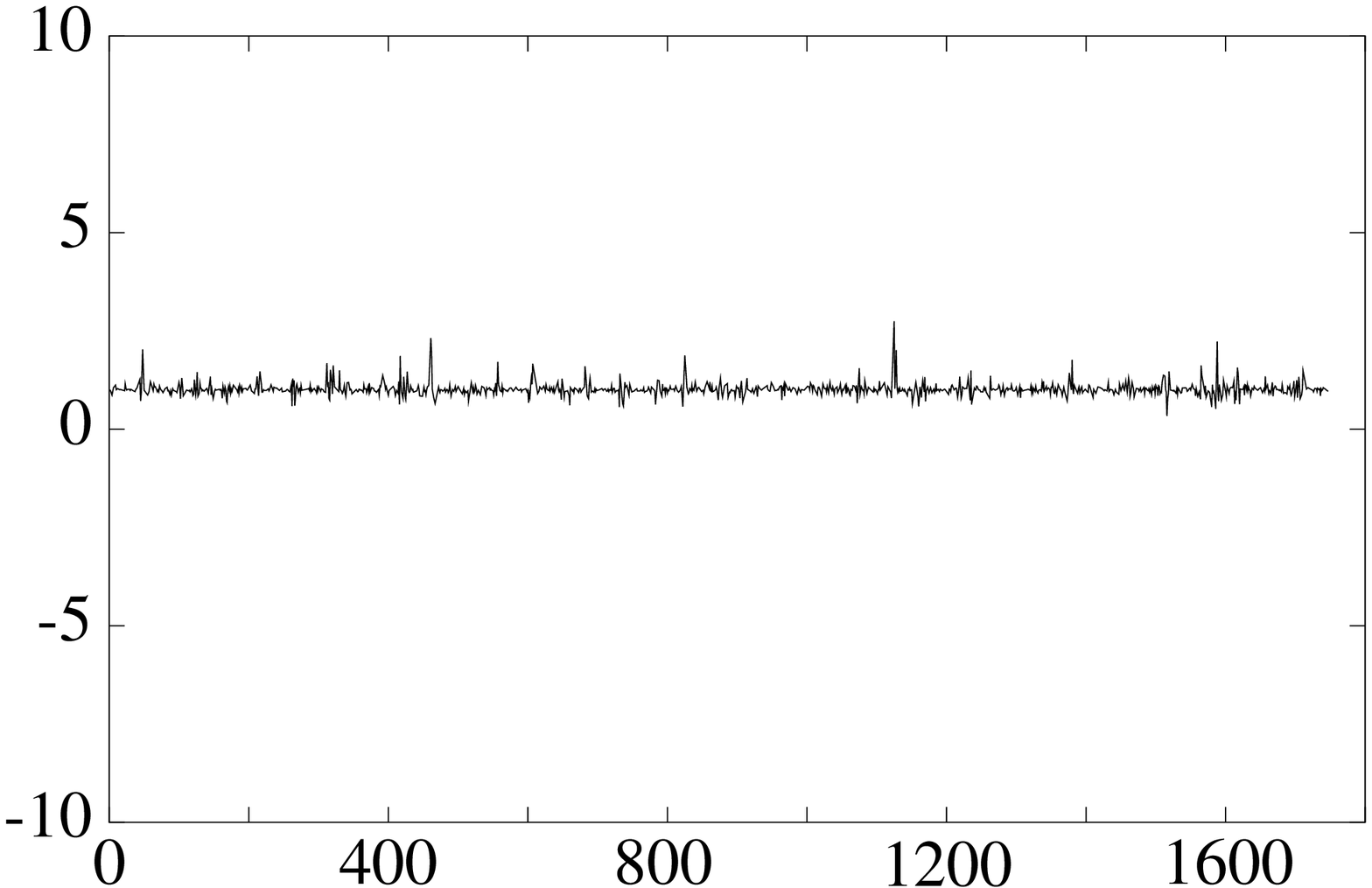}
\end{minipage}
\begin{minipage}{0.45\linewidth}
\includegraphics[width= \linewidth]{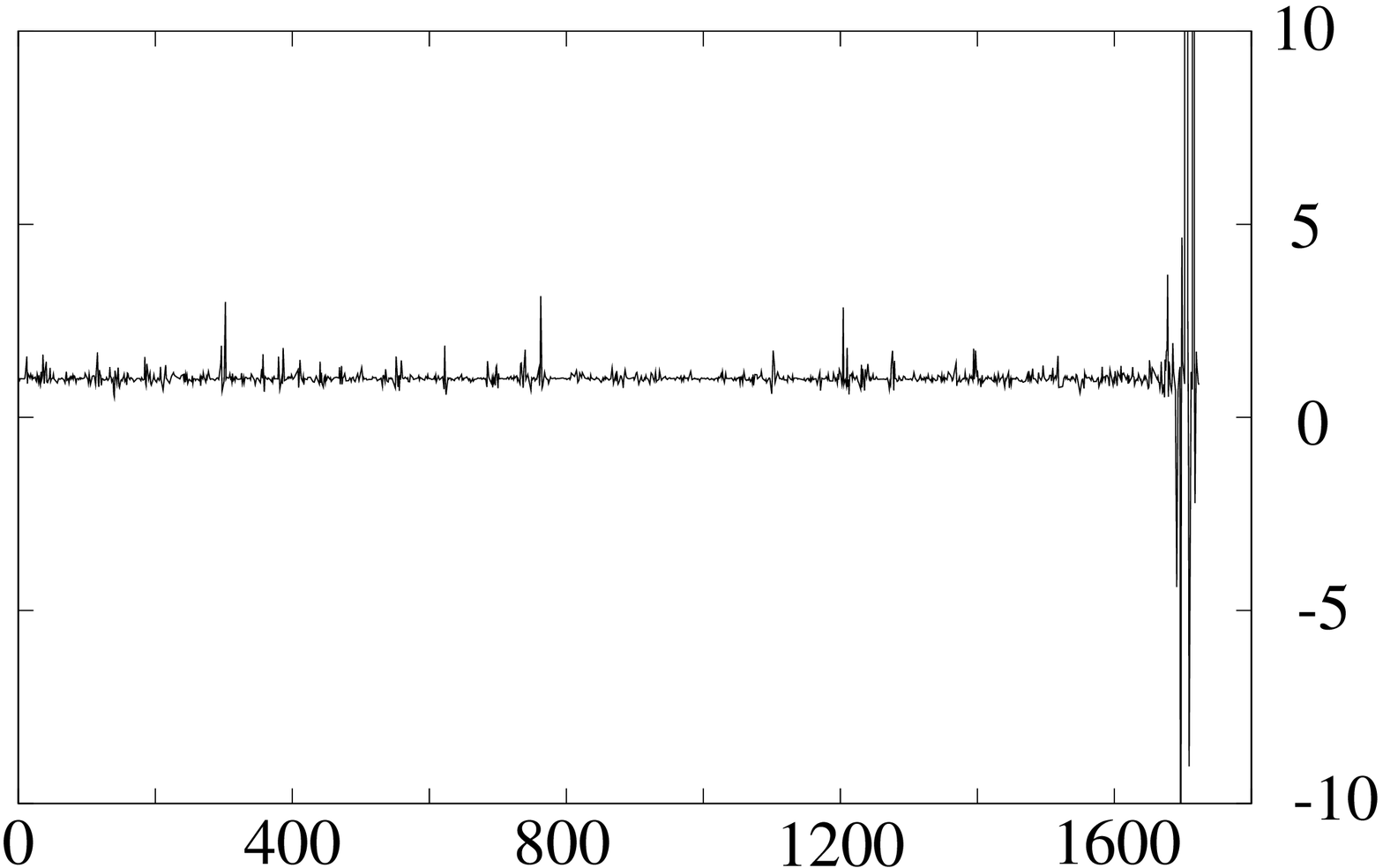}
\end{minipage}
\caption{The ratio of det$W$ as a function of
Monte Carlo sweeps. It fluctuates around one, but before
the simulation is crashed, big fluctuation is observed.}
\label{Fig-Det}
\end{figure}

The behavior of $<L>$ suggests that we are near the
phase transition.  But the calculation breaks because
of numerical instability at $\kappa=0.158$, $\mu=0.8$
and $\kappa=0.185$, $\mu=0.7$ in case of $4^4$, and we 
cannot go beyond.
To see origins of the instability, we measure
the ratio of the fermion determinant 
$\mbox{det} W(U+\Delta U)/\mbox{det} W(U)$.  Figure \ref{Fig-Det}
shows 
the behavior of the ration as a function of Monte
Carlo sweeps.  The ratio changes the value around one, but
suddenly it fluctuates very large and the calculation
is broken.  

\begin{figure}[htb]
\begin{minipage}{0.47\linewidth}
\includegraphics[width= \linewidth]{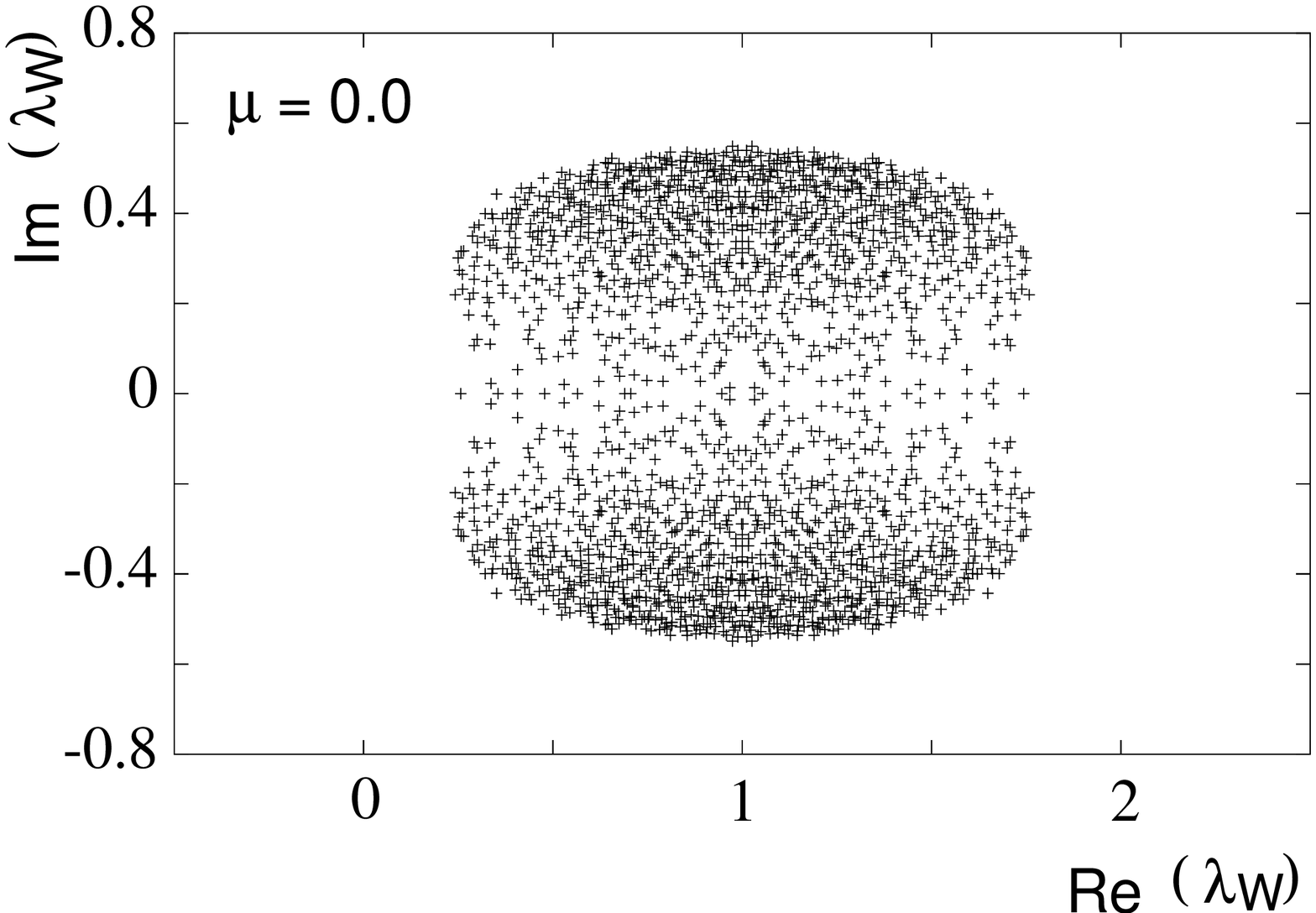}
\end{minipage}
\begin{minipage}{0.47\linewidth}
\includegraphics[width= \linewidth]{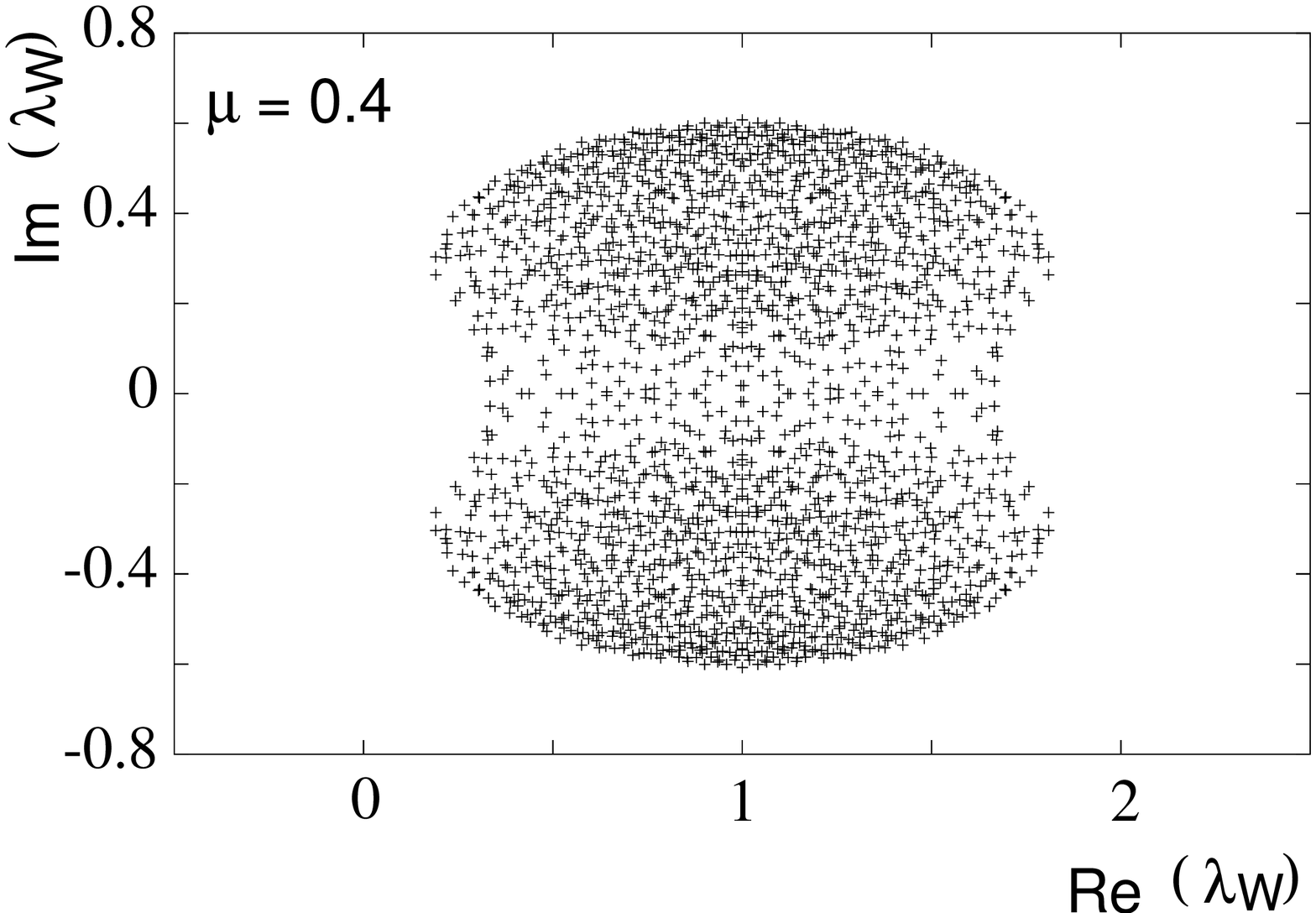}
\end{minipage}

\begin{minipage}{0.47\linewidth}
\includegraphics[width= \linewidth]{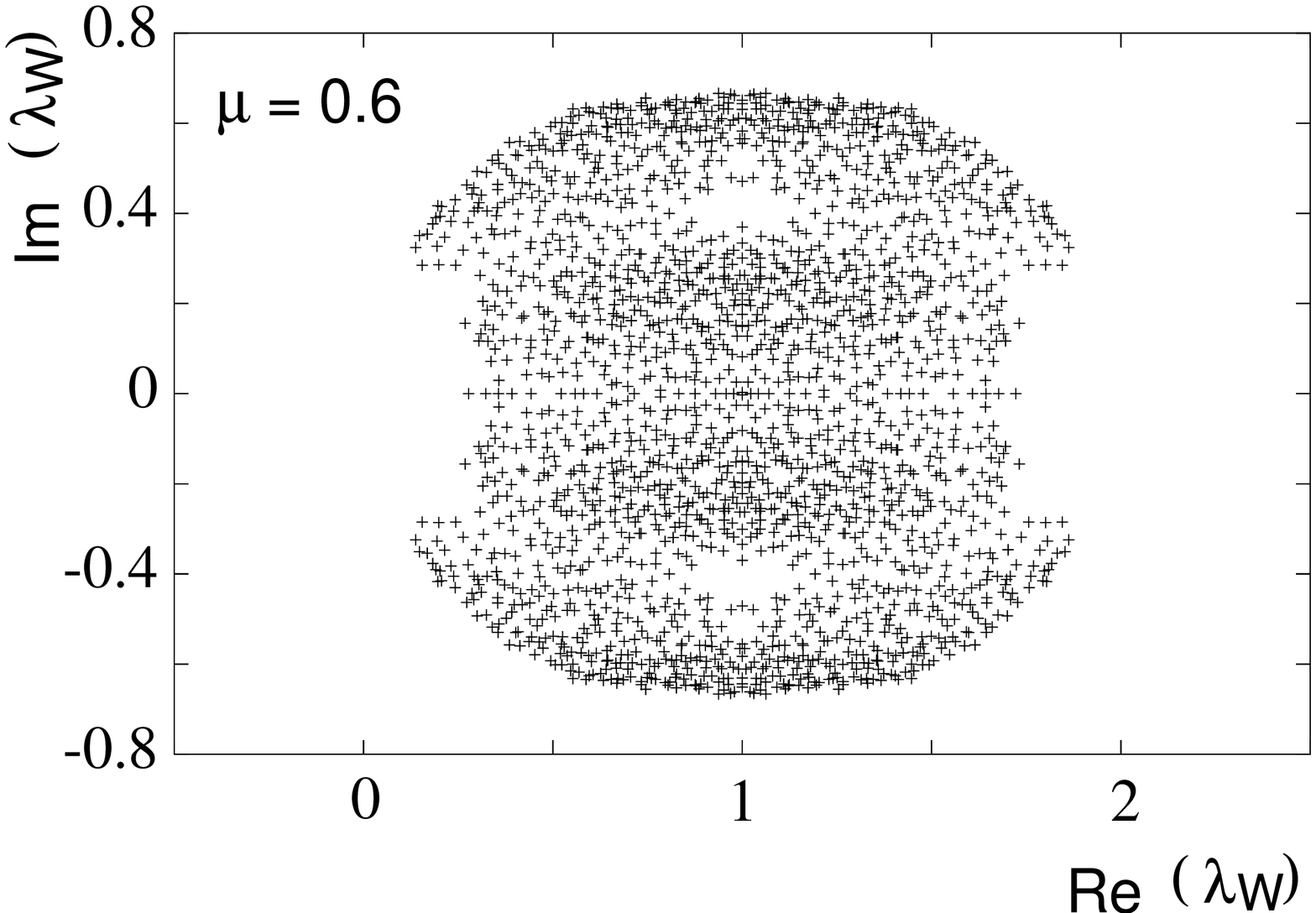}
\end{minipage}
\begin{minipage}{0.47\linewidth}
\includegraphics[width= \linewidth]{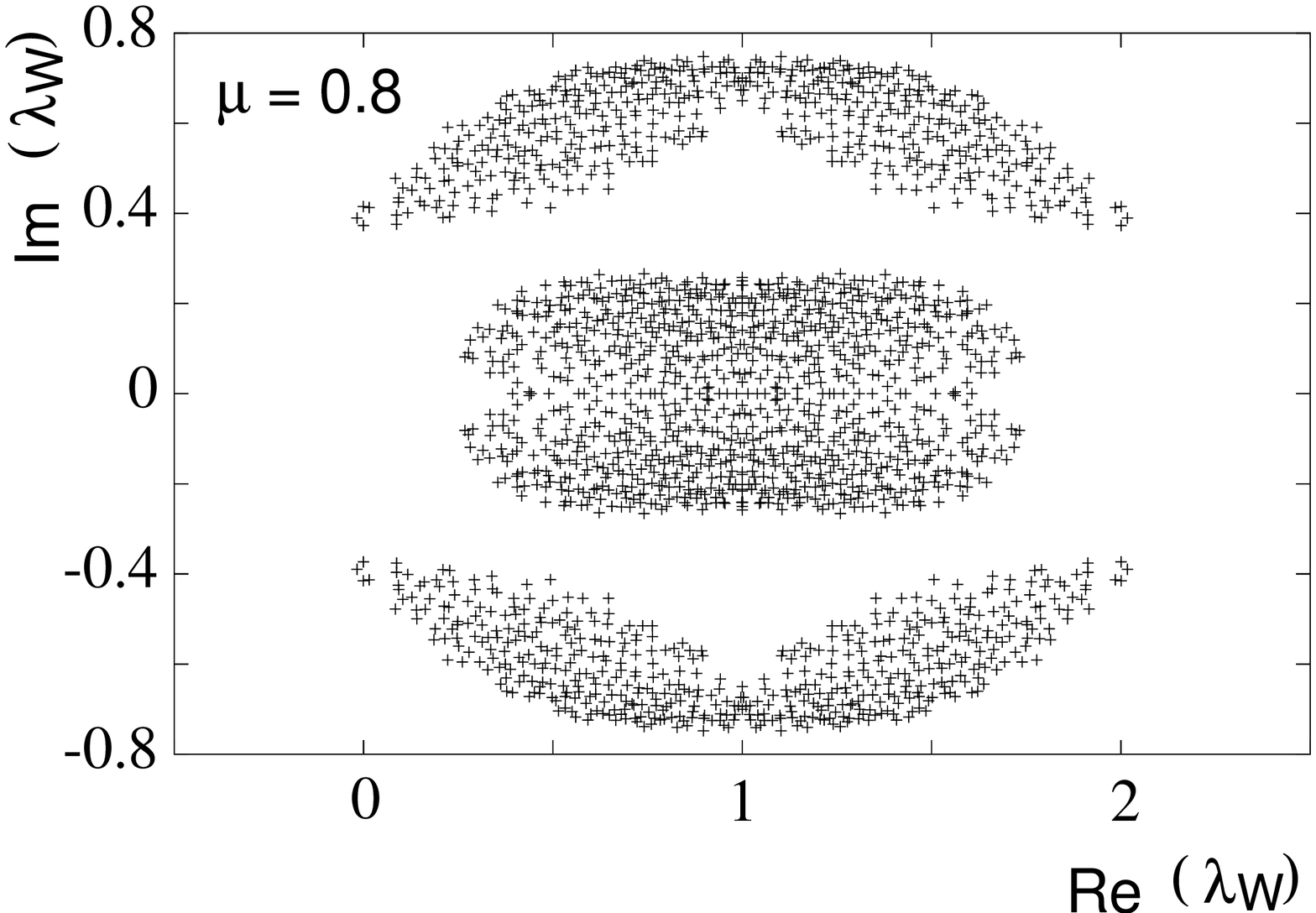}
\end{minipage}
\caption{Eigen value distribution of $W$ for $\mu$=0.0, 0.4,
0.6 and 0.8 at $\kappa=0.156$ on $4^4$ lattice.}
\label{Fig_Eigen}
\end{figure}

In Fig.\ref{Fig_Eigen}, we plot  eigen value distributions
of $W$ for $\mu=$0.0, 0.4, 0.6 and 0.8 at $\kappa=0.156$ on 
$4^4$ lattice.
As $\mu$ increases, the distribution of eigen values, $\lambda_i$,
becomes wide in the Real axis, and $\mbox{Min}\, \mbox{Re}(\lambda) < 0$; 
on the other hand, $\lambda_i$'s scatter sparsely around origin 
due to dynamical fermion simulation which includes det$W$ in the
measure.  

These behaviors are expected.  
But the behavior of the eigenvalues around the real axis
is new to us.  As $\mu$ deviates from zero, the region 
near the real axis becomes dilute, but it seems that a new
group appears when $\mu$ becomes further large, and
near the phase transition there are outer and inter
groups like the shell and bean.
This shell-and-bean structure occurs when the calculation
breaks, and might be related with the phase transition.

%%%%%%%%%%%%%%%%%%%%%%%%%%%%%%%%%%%%%%%%%%%%%%%%%
% eigen value of W (\kappa = 0.156))
% \begin{figure}
% \includegraphics[width= 0.7\linewidth]{pev21.eps}
% \caption{}
% \end{figure}
%%%%%%%%%%%%%%%%%%%%%%%%%%%%%%%%%%%%%%%%%%%%%%%%%
% eigen value of W (\kappa = 0.180)
% \begin{figure}
% \includegraphics[width= 0.7\linewidth]{pev22.eps}
% \caption{}
% \end{figure}
%%%%%%%%%%%%%%%%%%%%%%%%%%%%%%%%%%%%%%%%%%%%%%%%%
% eigenvalue of W (\kappa = 0.156)
%%%%%%%%%%%%%%%%%%%%%%%%%%%%%%%%%%%%%%%%%%%%%%%%%%%%%
% eigen vale of W (\kappa = 0.185)
% \begin{figure}
% \begin{minipage}{0.5\linewidth}
% \includegraphics[width= \linewidth]{pevm00k185.eps}
% \caption{det $W$}
% \end{minipage}
% \hspace{0.5cm}
% \begin{minipage}{0.5\linewidth}
% \includegraphics[width= \linewidth]{pevm40k185.eps}
% \caption{det $W$}
% \end{minipage}
% \end{figure}
%
% \begin{figure}
% \begin{minipage}{0.5\linewidth}
% \includegraphics[width= \linewidth]{pevm60k185.eps}
% \caption{det $W$}
% \end{minipage}
% \hspace{0.5cm}
% \begin{minipage}{0.5\linewidth}
% \includegraphics[width= \linewidth]{pevm70k185.eps}
% \caption{det $W$}
% \end{minipage}
% \end{figure}

%%%%%%%%%%%%%%%%%%%%%%%%%%%%%%%%%%%%%%%%%%%%%%%%%%%%%%
\begin{table}[htb]
\caption{Pseudo scalar and vector masses at
$\mu$=0.0, 0.2, 0.4 and 0.5 for $\kappa$=0.158 and
0.180.
}
\begin{center}
\begin{tabular}{|c||c|c|}
\hline 
 & \multicolumn{2}{|c|}{$\pi $} \\  
\hline
$\mu $ & $\kappa = 0.158 $ & $\kappa = 0.180$  \\
\hline
0.0 &1.775(31)  & 1.327(51)  \\
\hline 
0.2 &1.722(09) & 1.305(37)  \\
\hline 
0.4 &1.738(11) & 1.428(51)  \\
\hline 
0.5 & 1.649(18) & 1.346(65)  \\
\hline
\hline
& \multicolumn{2}{c|}{$\rho $} \\
\hline
0.0  & 1.882(45) & 1.509(41) \\
\hline 
0.2 &  1.783(13) & 1.490(41) \\
\hline 
0.4 & 1.788(36) & 1.660(49) \\
\hline 
0.5 &  1.798(46) & 1.482(100) \\
\hline
\end{tabular}
\end{center}
\end{table}

\begin{figure}
\begin{center}
\includegraphics[width= 0.9\linewidth]{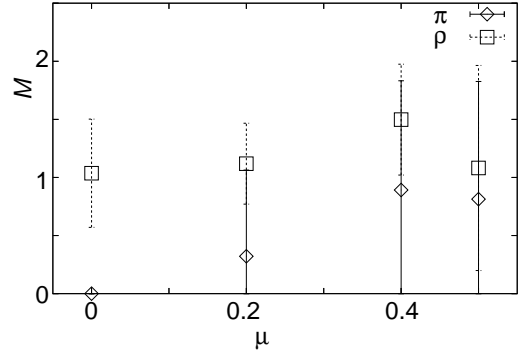}
\end{center}
\caption{$\pi$ and $\rho$ masses as a function of the chemical
potential $\mu$ after extrapolation to the chiral limit.}
\label{Fig-Mass}
\end{figure}

We evaluate pseudo scalar and vector meson masses
for $\mu=$0.0, 0.2, 0.4 and 0.5.  Because the lattice
is small, we fit propagators at $N_t=$2,3,4,5 and 6
by one-pole fit.  The result is shown in Table 1.
Although the data are still very preliminary, we
extrapolate them to the chiral limit obtain Fig.\ref{Fig-Mass}.
Error bars are very large, but the pion mass becomes massive
around $\mu \sim 0.4$, which means that
the chiral symmetry is restored in these regions. 

%%%%%%%%%%%%%%%%%%%%%%%%%%%%%%%%%%%%%%%%%%%%%%%%%%%%%
% Mass 
%%%%%%%%%%%%%%%%%%%%%%%%%%%%%%%%%%%%%%%%%%%%%%%%%%%%%

%%%%%%%%%%%%%%%%%%%%%%%%%%%%%%%%%%%%%%%%%%%%%%%%%%%%%%

\section{CONCLUDING REMARKS}

We present numerical study of two-color QCD with the
chemical potential with Wilson fermions at $\beta=1.6$.
Although the lattice is very small, most data suggests
we are reaching the confinement/deconfinement phase
transition.

We employ an algorithm which takes into account
the ratio of fermion determinant exactly, and has large
Markov step, but we suffer from numerical instability and can
not go over the phase transition.  Near the phase transition,
the distribution of eigen values of $W$ shows a peculiar
``Shell-and-Bean'' structure.

Since the calculation is done at strong coupling region, the
strange behavior of $\mbox{det} W$ might be related with rough
configurations far from the continuum \cite{Stamatescu}.
We plan to continue the analysis by using improved gauge
actions to clarify the point.

\subsection*{Acknowledgment}

Numerical simulations have been done at RCNP at Osaka university
and
Science Information Processing Center at Tsukuba university.
Grant-in-Aid
This work is supported by the Grant-in-Aide for
Scientific Research by JSPS and Monbusho, Japan 
(No. 10640272 and No. 12554008).
One of the authors (A.N.) thanks M-L.Lombardo, 
M.A.Stephanov and J.J.M.Verbaarschot for useful discussions
and staffs at INS of Univ. Washington for their hospitality
where the work started.

%%%%%%%%%%%%%%%%%%%%%%%%%%%%%%%%%%%%%%%%%%%%%%%%%%%%%%

\end{document}